\documentstyle[prd,aps,epsf]{revtex}

\begin{document}
\draft
\author{S.~W.~Mansour and T.~K.~Kuo}
\address{Department of Physics, Purdue University, West Lafayette, Indiana 47907}
\title{Solar Neutrinos and the Violation of Equivalence Principle}
\date{}
\maketitle 
\begin{abstract}
In this Brief Report, a non-standard solution to the solar neutrino problem is revisited. This solution assumes that neutrino flavors could have different couplings to gravity, hence, the equivalence principle is violated in this mechanism. The gravity induced mixing has the potential of accounting for the current solar neutrino data from several experiments even for massless neutrinos. We fit this solution to the total rate of neutrino events in the SuperKamiokande detector together with the total rate from other detectors and also with the most recent results of the SuperKamiokande results for the recoil-electron spectrum.
\end{abstract}
\pacs{}
\section{Introduction}
At present several neutrino detectors indicate that there is a major discrepancy between the Standard Solar Model (SSM) predictions for the solar neutrino flux~\cite{a:1} and its observed value. Recently, five neutrino experiments (GALLEX~\cite{a:2}, SAGE~\cite{a:3}, Homestake~\cite{a:4}, Kamiokande~\cite{a:5} and SuperKamiokande~\cite{a:6}) reported that the observed neutrino flux is less than the values predicted by the SSM for different neutrino energies. 
The first three experiments detect the solar electron neutrinos, $\nu_e$, via absorption processes so they do not give any information about the direction and energy spectrum of the observed neutrino events. The last two water Cherenkov detectors employ $\nu_e$-$e$ scattering so the direction and energy distribution of the detected neutrino can be fairly determined. This would allow us to test theoretical models not only against the measured total rate but also against the energy spectrum of the neutrino flux.
The analysis of the first 504 days data from SuperKamiokande~\cite{a:7}(SK), when combined with the data from earlier experiments, provide us with important constraints on the famous MSW effect~\cite{a:8} and vacuum oscillation solutions of the solar neutrino problem. The MSW effect involves the standard neutrino interactions with the matter background in the Sun, leading to an enhanced $\nu_e$-$\nu_x$ resonant transition. 
The ``just so'' vacuum oscillation solution explains the suppression using only vacuum oscillations of the neutrinos on their way to the Earth.
This solution is in fact in slightly better agreement with the recent data from SK than the matter-induced MSW effect~\cite{a:9}.  However, it requires some ``fine tuning'' of the Sun-Earth distance to account for the apparent deficit.
Both of these solutions require that neutrinos have non-degenerate masses; they will be referred to as the Mass-Induced (MI) scenario.  
 
In this report, we explore yet another possible solution of the solar neutrino problem. However, it requires a rather unorthodox assumption that the neutrino flavors couple differently to gravity. This is a statement of the violation of the weak equivalence principle which stipulates that all matter couple equally to gravity. 
This solution has been suggested by different authors~\cite{a:10,a:11}; here, we consider the Gravitational-Induced (GI) scenario in the light of the most recent results of the total rate and recoil-electron spectral shape from the SK detector. 
It should be noted that the pure GI mechanism does not require the introduction of neutrino masses and thus can be consistent with a zero or degenerate mass scheme for light neutrinos ($\nu_e$, or $\nu_{\mu}$). 
We use a two-flavor analysis treatment in our analysis for simplicity. A complete GI three-flavor analysis treatment was studied~\cite{a:12}; however, the present experimental results are not enough to constrain the parameter space for a three-flavor analysis. So the assumption made here is that the GI mechanism, if it were to be the cause of the solar neutrino problem, is due to only two neutrino flavors.  
A mixed GI+MI scenario where neutrinos are assumed to be massive was studied by the authors of~\cite{a:13} assuming that the MI effect is the main cause of the neutrino flux deficit. We do not cover this possibility here.

In addition to the GI scenario considered here, there are other scenarios which are related to the violation of equivalence principle. The first involves the neutrino coupling to the massless dilaton field which arises from String Theory~\cite{a:14,a:15}. This scenario allows for a vacuum oscillation solution of the solar neutrino problem, but there is no clear distinction to be made in the present neutrino experiments between it and the conventional mass mixing.
The second one proposed by Glashow et al.~\cite{a:16} assumes that Lorentz invariance is violated during neutrino propagation. The analysis is equivalent to the GI scenario with a redefinition of parameters. In this scenario, different neutrino flavors are allowed to have different velocities under the assumption that the gravitational potential $\Phi(r)$ does not change appreciably over the distance of interest.

The article is organized as follows. Section II reviews the 2-flavor formalism for the pure GI mechanism of neutrino oscillations, detailing the survival probability of the neutrinos during their propagation in a gravitational potential. Section III presents a $\chi^2$ analysis of the solar neutrino data which is used to constrain the relevant GI parameters. Some concluding remarks are given in Section IV. 

\section{GI Formalism for Two Neutrino Flavors}

The violation of the equivalence principle has been tested in several famous experiments by Eotvos, Dicke et al.~, etc.~\cite{a:17}. The current bounds on the extent of the violation still allow an appreciable parameter space for the GI mechanism to have an effect on the neutrino propagation in regions where we have large gravitational fields.
We discuss the case of a pure GI mechanism where the neutrinos are assumed to be massless. The neutrino wavefunction $\psi^i(x)$, where $i$ denotes its flavor, obeys the gravity modified Dirac equation,
$$ \left[\eta^{\mu\nu}+h_i^{\mu\nu}(x)\right]\partial_\mu\partial_\nu \psi^i(x)= 0,\eqno(2.1)$$
where $h_i^{\mu\nu}(x)$ are the perturbations in the metric due to the gravitational potential $\Phi(x)$. The $h_i^{\mu\nu}$'s can be parametrized as $f_i G^{\mu\nu}$, where $f_i$ are the flavor coupling constants to the gravitational field $G^{\mu\nu}$. 
The constants $f_i$ can be regarded as post-Newtonian (PPN) parameters.
If Einstein's equivalence principle is not obeyed, the parameters $f_i$ will not be equal for different neutrino flavors.

In the GI mechanism the neutrino weak basis $\nu_\alpha$ does not coincide with the gravity basis $\nu_a$ and the neutrino eigenstates mix according to
$$\nu_\alpha = U^G_{\alpha a}\nu_a,\eqno(2.2)$$
where $U^G$ is a unitary matrix depending on the gravity mixing angle $\theta_G$.
The correspondence between the GI and the MI formalisms is obvious if we perform the following replacements: 
$$\theta_G \rightarrow \theta_{MI}$$ and 
$$2 E \delta f \Phi \rightarrow {\delta m^2\over 2E}.\eqno(2.3)$$
$\delta f\equiv f_e-f_x$ denotes the difference between the gravitational couplings of $\nu_e$ and $\nu_x$ where we will be considering the transition between the electron-neutrino flavor $\nu_e$ and another flavor $\nu_x$ ($\nu_\mu$, or $\nu_\tau$) which is relevant to the solar neutrino problem. $\delta m^2$ is the corresponding squared mass difference in the MI effect.

In the presence of ordinary matter, the propagation equation for {\em massless} neutrinos in a gravitational field can be written as 
$$i {d\over dx} \left(\matrix{\nu_e \cr \nu_x}\right)=2 E  \delta \gamma \left(\matrix{A & {1\over 2}\sin 2 \theta_G \cr {1\over 2}\sin 2\theta_G & \cos 2\theta_G} \right) \left(\matrix{\nu_e \cr \nu_x}\right),\eqno(2.4)$$
where $E$ is the neutrino energy and $\delta \gamma \equiv \left|\Phi\right| \delta f$. We shall use $\delta \gamma$ in our analysis instead of $\delta f$ due to the ambiguity in choosing the potential $\Phi(r)$~\cite{a:11}\footnote{Many authors suggest that the correct choice of the potential lies in choosing the potential of the Great Attractor, $\Phi\sim 10^{-5}$, not the solar potential, $\Phi\sim 10^{-6}$. We will assume the former potential to be the dominant in our analysis especially that it has no considerable variation over the relevant distance scale.}.
$A\delta \gamma \equiv \sqrt{2} G_F N_e/ 2 E$ represents the charged weak interaction term between the neutrinos and the electrons in matter with a number density $N_e$. 
In vacuum, the survival probability can be easily obtained,
$$P(\nu_e \rightarrow \nu_e)=1 - \sin^2 2\theta_G \sin^2 {\pi l \over \lambda_G},\eqno(2.5)$$
where $\lambda_G$ is the GI oscillation length defined as
$$\lambda_G={\pi \over E \left|\Phi\right| \delta f}.\eqno(2.6)$$
The energy dependence of $\lambda_G$ is obviously different from the case of the MI oscillation length which is inversely proportional to $E$. This has led several authors~\cite{a:11} to suggest that the best way to distinguish between the two mechanisms is by measuring the neutrino energy spectrum. 

Now we turn to the generalization of Eq.~(2.5) in the presence of matter. By looking at the propagation equation, Eq.~(2.4), we note that we can have a resonant transition between $\nu_e$ and $\nu_x$ only if $\delta f$ is positive and
$$2 E \delta \gamma = {\sqrt{2} G_F N_e(r) \over \cos 2 \theta_G}.\eqno(2.7)$$
The adiabaticity condition in the GI mechanism reads 
$$\gamma\equiv{\delta \gamma E \sin^2 2 \theta\over \left|{1\over N_e}{dN_e\over dr} - {1\over\Phi} {d\Phi\over dr}\right|_{\rm resonance}\cos 2 \theta_G} \gg 1,\eqno(2.8)$$ 
where we assume in our calculation that the variation of the gravitational potential of the Great Attractor with distance is slow, i.~e.~ ${1\over\Phi}{d\Phi\over dr} \ll {1\over N_e}{dN_e\over dr}$. 
Using the analogy to the derivation of the MSW survival probability, we can write the survival probability in the presence of matter as
$$P(\nu_e \rightarrow \nu_e)={1\over 2}+\left({1\over 2}-P_c\right)\cos 2\theta_G \cos 2\theta_{Gm},\eqno(2.9)$$
where the matter mixing angle $\theta_{Gm}$ is defined as, $\tan 2\theta_{Gm}={\sin 2\theta_G /(\cos 2\theta_G - A)}$ and $P_c$ is the level-crossing probability given by its standard form for an exponentially varying density,
\begin{eqnarray}
P_c&=&(\exp[-{\pi\over 2} \gamma (1-\tan^2 \theta_G)]-\exp[-2\pi \gamma \cos 2 \theta_G/\sin^2 2\theta_G])/\nonumber\\ & &(1-\exp[-2\pi \gamma \cos 2 \theta_G/\sin^2 2\theta_G])\nonumber.\\
\eqnum{2.10}
\end{eqnarray}
The survival probability is plotted versus $E \delta \gamma$ for different values of $\sin^2 2\theta_G$  in Fig.~\ref{fig;prob}. 

\section{Numerical Analysis}

In the following discussion, we present a $\chi^2$-analysis of the most recent solar neutrino data in the light of the GI mechanism. In Section IIIa, a $\chi^2$ analysis on the total rate of the neutrino flux is first performed to obtain the constraints on the GI parameters ($\delta \gamma$, $\sin^2 2 \theta_G$). The following section, Section IIIb, deals with the analysis of the recoil-electron spectrum from the SK detector. Both analyses are then combined to constrain the relevant GI parameters.   
\subsection{Total Rates}
The experimental results of the total rates for the four neutrino experiments used in this article are shown in Table~\ref{table;exprates}. We use the predictions of the Bahcall-Pinsonneault standard solar model of Ref.~\cite{a:18} (BP98) assuming the Institute of Nuclear Theory (INT) estimate~\cite{a:19} for the B$^8$ production cross section.
Our procedure takes into account the theoretical uncertainties in the theoretical estimates of the neutrino flux. The $\chi^2$ function is given by~\cite{a:9} 
$$\chi^2_{\rm{rates}}=\sum_{i,j=1,\ldots, 4}\left(R_i^{\rm th}-R_i^{\rm exp}\right) V_{i j}^{-1} \left(R_j^{\rm th}-R_j^{\rm exp}\right),\eqno(3.1)$$
where $R_i^{\rm th (exp)}$ represents the GI theoretical predictions (experimental values) of the neutrino detection rate divided by the SSM predictions for the $i$-th experiment. $V_{i j}$ is the error matrix which contains the theoretical uncertainties as well as the experimental errors for each experiment.
Fig.~\ref{fig;rates} shows the 90\% and 95\% C.~L.~regions for the GI parameters $\delta \gamma$ and $\sin^2 2\theta_G$ with $\chi^2_{\rm min}=0.46$. 
This is to be compared to our estimate of the MI best fit corresponding to the MSW small angle solution to the solar neutrino problem, where $\chi^2_{\rm min}=0.62$. 
In Fig.~\ref{fig;rates}, we see that there are two allowed regions: the first is at $\delta \gamma \sim 10^{-18}$ and $\sin^2 2 \theta_G \sim 10^{-3}$ (small angle solution);
while the second region appears at a smaller value of $\delta \gamma \sim 10^{-21}$ and maximal mixing where $\theta_G\lesssim \pi/4$ (large angle solution).
The analogy between the standard MI solutions and the GI solutions is understood since the GI calculations involved with the rate analysis come straight from the MI calculations with the parameter replacement of Eq.~(2.3).
\subsection{Recoil electron spectrum}
\begin{figure}
\caption{Contour plot of the exclusion region in the $(\delta \gamma, \sin^2 2\theta)$ plane, the dotted curve shown is an isocontour at $\chi^2=25.0$ for 14 D.~O.~F.~ This has been based on the recent results of the recoil-electron spectral shape from the 504 days SK data run.
The regions allowed from the rate analysis are also shown. As in Fig.~1, the dashed curve represents the 90\% C.~L.~while the solid curve represents the 95\% C.~L.~} \label{fig;spec2}
\end{figure}
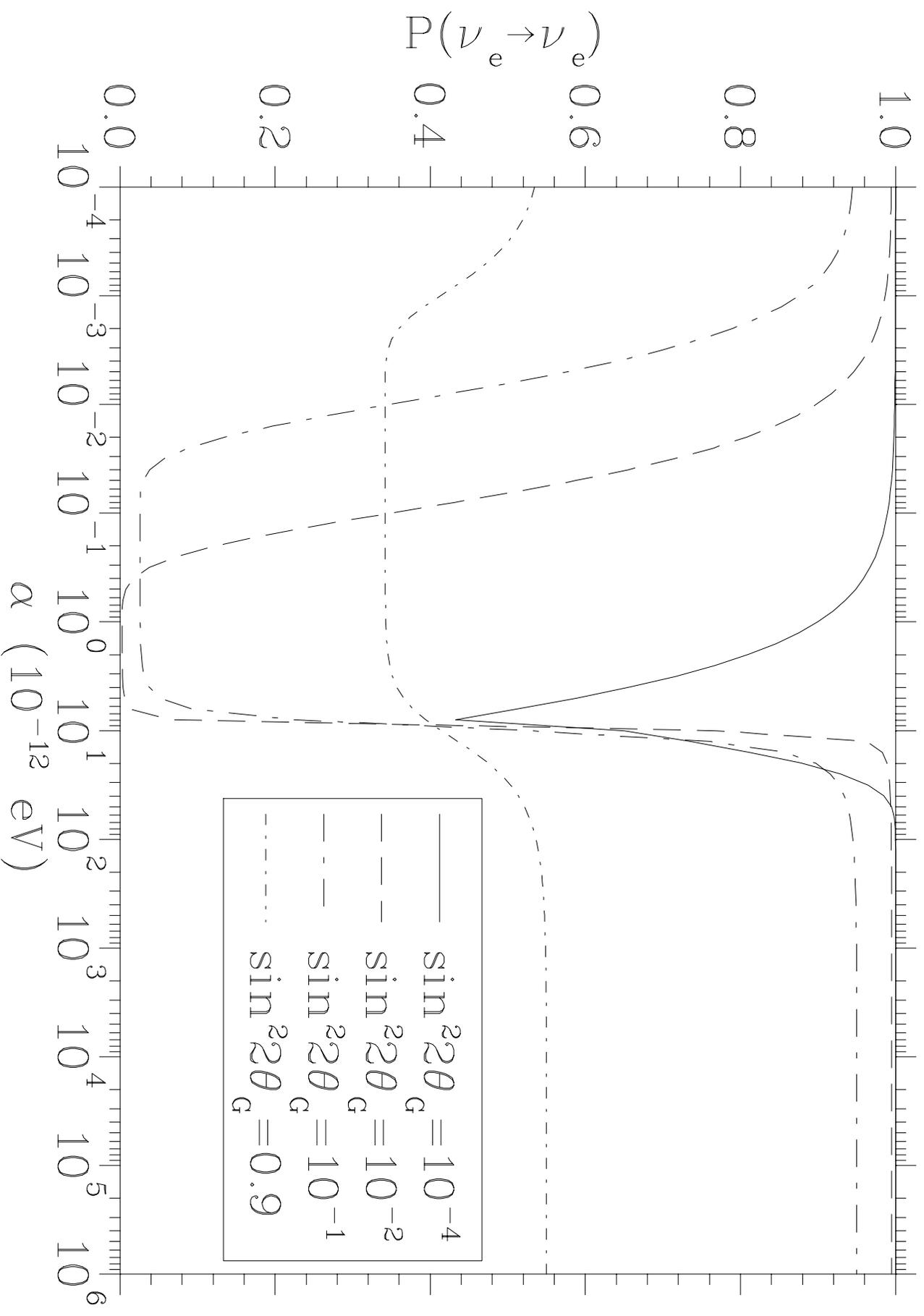
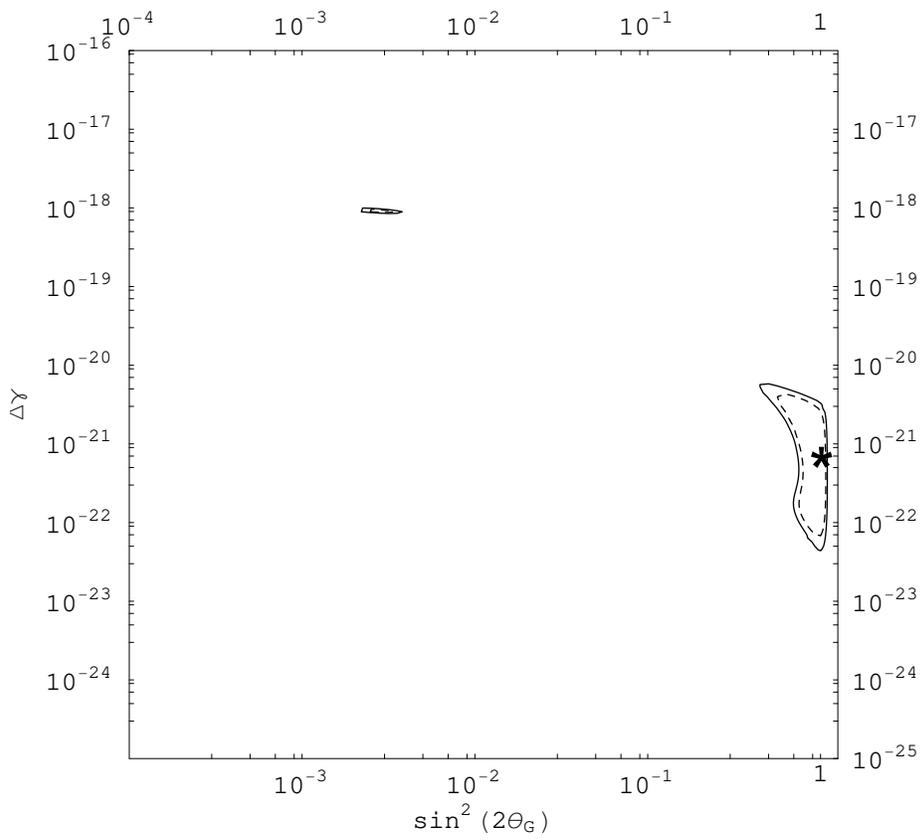
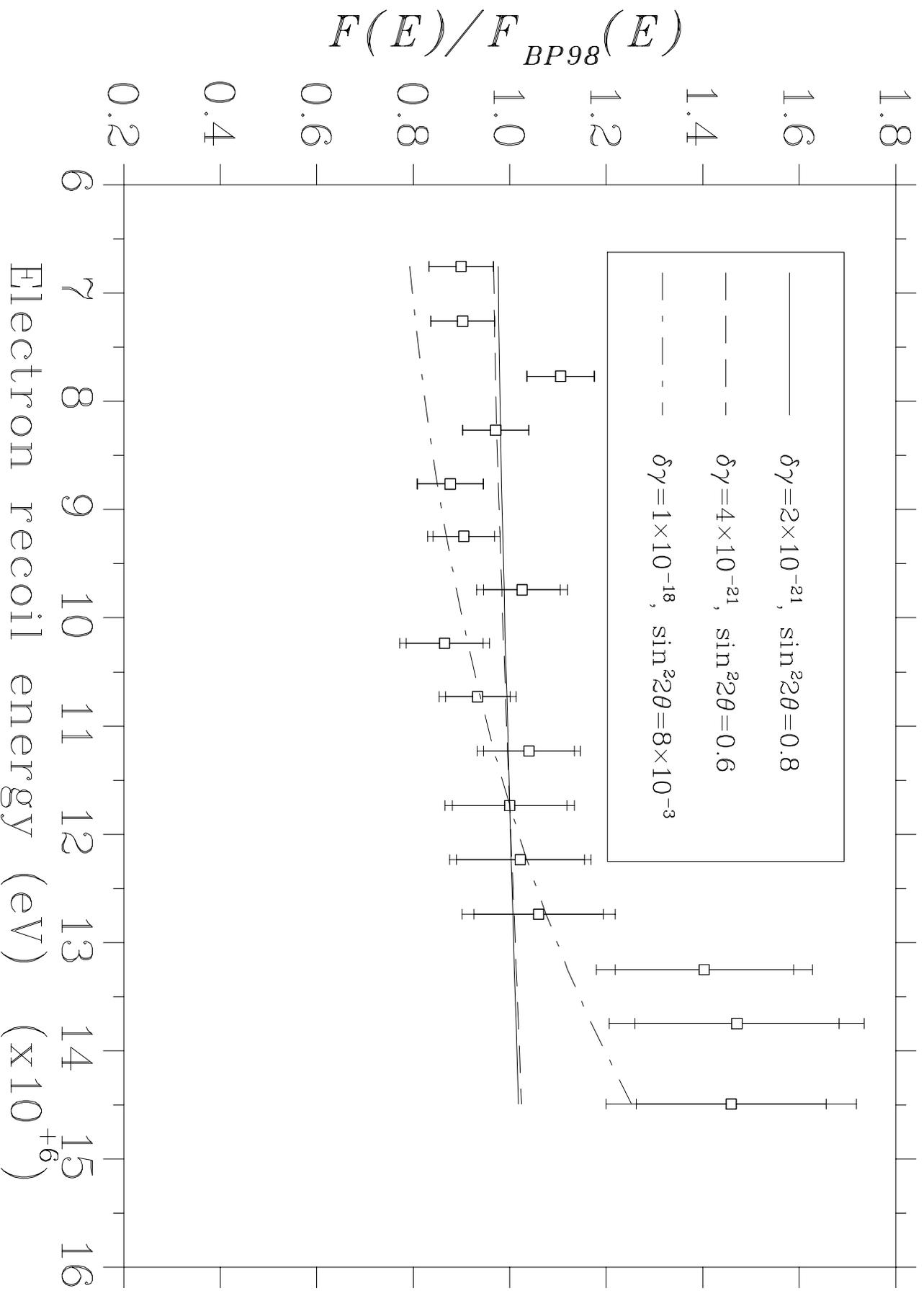
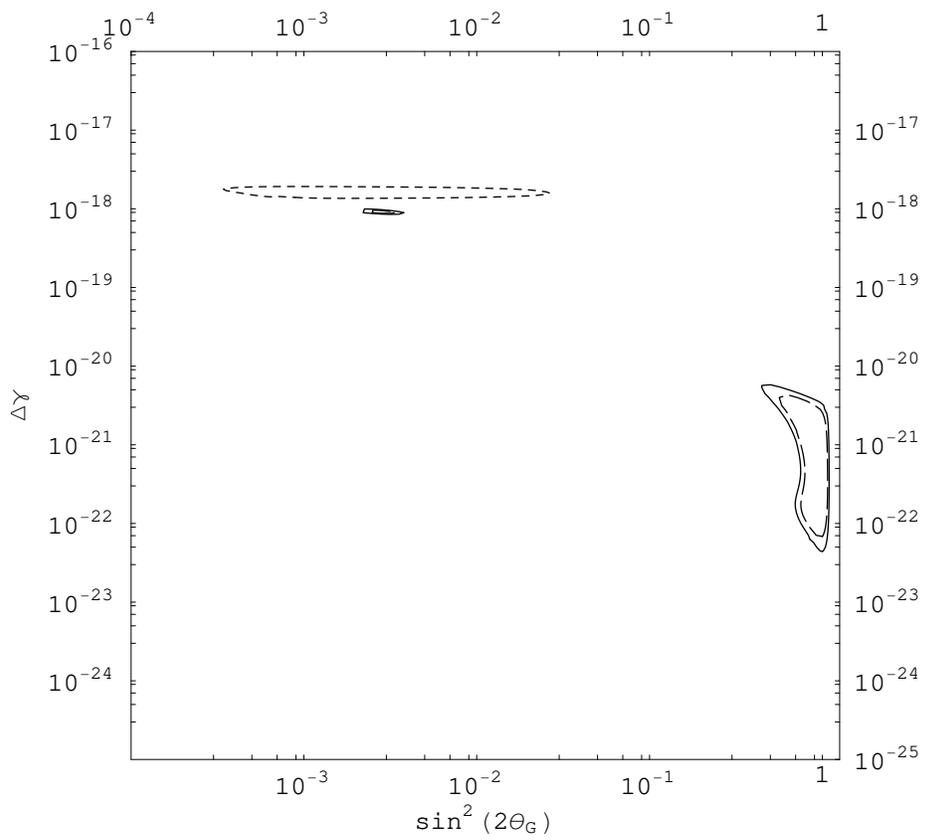

We next consider the analysis of the SK recoil electron spectrum using the following $\chi^2$:
$$\chi^2_{\rm spectra}=\sum_{i,j=1, \ldots, 16}\left(\alpha \phi_i^{\rm th} - \phi_i^{\rm exp} \right) V'^{-1}_{i j} \left(\alpha \phi_j^{\rm th}-\phi_j^{\rm exp}\right).\eqno(3.2)$$ 
Here $\phi_i^{\rm th (exp)}$ is the $i$-th energy bin theoretical (experimental) value for the measured flux divided by the SSM prediction. 
The error matrix used here is given as $V'_{ij}\equiv \sigma^{\rm stat}_i \sigma^{\rm stat}_j\delta_{ij} + \sigma^{\rm sys}_i \sigma^{\rm sys}_j$~\cite{a:9}.
$\alpha$ is the flux normalization parameter which is allowed to vary independently of $\delta \gamma$ and $\sin^2 2 \theta_G$. This variation allow for the testing of measured spectrum and not the overall rate of SK.
Fig.~\ref{fig;spec1} shows the ratio of the GI predicted flux to the SSM predictions for the $^8$B neutrino flux plotted against the recoil electron energy. The predicted ratio and the experimental results are normalized at 11.7 MeV.
Fig.~\ref{fig;spec2} presents the spectral shape analysis exclusion region at the $\chi^2=25.0$ level for 14 degrees of freedom (16 energy bins minus two free parameters) together with the allowed regions from the rate analysis. We note that both the small angle solution and the large angle solution from the rate analysis are not affected. 

In the spectral shape analysis, we allow the hep flux contribution to vary as suggested in Ref.~\cite{a:20} to see whether we will get any effect on the exclusion region or not. It is found that an hep flux enhancement of about $10^3$ is needed to have any appreciable effect on the exclusion region. In any case, the exclusion region does not intersect with any of our allowed regions from the rate analysis. Thus, the spectral shape analysis based on the recent SK data yields no additional constraints on the GI allowed parameter space. 
 
\section{Discussion}
We have shown in this report that the GI scenario is still a viable solution to the solar neutrino problem. 
From just the rate analysis, we obtain two allowed regions in the GI parameter space: the first is at $\delta \gamma \sim 10^{-21}$ and maximal mixing, while the second is at a smaller value of $\delta \gamma (\sim 10^{-18})$ and $\sin^2 2\theta \sim 10^{-3}$. The spectral shape analysis has no effect on these allowed regions.
We should also note that the combined best fit regions are still allowed by the current experimental bounds on the violation of the equivalence principle and by the neutrino accelerator experiments~\cite{a:21}. 
Although the allowed parameter regions for the GI scenario are quite small and may seem unlikely, they are by no means excluded by the available data.
Thus, to obtain the final word about either the MI or GI explanations to the solar neutrino problem, more data are needed especially on the solar neutrino spectral shape from the SK and SNO detectors so that we can eventually accept or reject one of these scenarios.

\acknowledgments 
We are grateful to Jim Pantaleone and Terry Leung for useful discussions. S.~M.~is supported by a research grant from the Purdue Research Foundation. T.~K.~is supported by the Department of Energy Grant No.~DE-FG02-91ER40681B.

\pagebreak
\newpage
\newpage
\widetext
\begin{table}
\caption{The current experimental results and SSM predictions for the total rate in four neutrino detection experiments. The rate is measured in SNU for all of the experiments except the SuperKamiokande detector. The SSM predicted rates are based on the Bahcall-Pinsonneault standard solar model (BP98) of Ref.~[18]. The theoretical errors are at the 1$\sigma$ level.} \label{table;exprates}
\begin{tabular}{@{\hspace{.4in}}llll@{\hspace{.4in}}}
Experiment & Experimental Rate & SSM predicted Rate & Threshold Energy\\
\hline
Homestake (Cl)~\cite{a:4} & $2.56\pm 0.16 \pm 0.14 $ & $7.7 \begin{array}{l}+1.2\\-1.0 \end{array}$& 0.81 MeV \\
GALLEX (Ga)~\cite{a:2} & $77.5 \pm 6.2 \begin{array}{l}+4.3\\-4.7 \end{array}$ & $ 129 \begin{array}{l}+8\\-6 \end{array}$ & 0.24 MeV\\
SAGE (Ga)~\cite{a:3} & $66.6 \begin{array}{l}+7.8\\-8.1 \end{array}$ & $129\begin{array}{l}+8\\-6 \end{array}$ & 0.24 MeV\\ 
SuperKamiokande\footnote{The SuperKamiokande rate is measured in $10^6$ cm$^{-2}$s$^{-1}$.}~\cite{a:6} & $2.44 \pm +0.05\begin{array}{l}+0.09\\-0.07 \end{array}$ & $5.15 \begin{array}{l}+1.0\\-0.7 \end{array}$ & 6.5 MeV\\
\end{tabular}
\end{table}
\begin{figure}
\caption{The survival probability $P(\nu_e \rightarrow \nu_e)$ in the GI scenario. It is plotted against the product of the neutrino energy $E$ and the measure of the equivalence principle violation $\delta \gamma$ ($\alpha=E\delta \gamma$ in units of $10^{-12}$ eV).}
\label{fig;prob}
\end{figure}
\begin{figure}
\caption{Contour plot of the allowed regions in the $(\delta \gamma, \sin^2 2\theta)$ plane.
These are found from the results of the total rate for the four neutrino experiments (HomeStake, GALLEX, SAGE and SuperKamiokande). The GALLEX and SAGE results have been combined in the analysis. The solid curve represents the 95\% C.~L.~region while the dashed one indicates the 90\% C.~L.~region. The star shows the position of the best fit solution at $\chi^2_{min}$.}\label{fig;rates} 
\end{figure}
\begin{figure}
\caption{The ratio of the GI predicted recoil-electron spectrum $F(E_e)$ to the SSM spectrum $F_{SSM}(E_e)$ for different values of $\delta \gamma$ and $\theta_G$. The ratio is normalized at $E_e=11.7$ MeV. The error bars indicate the statistical and systematic errors of the experimental data.}
\label{fig;spec1}
\end{figure}
\end{document}